# ADDiff: Semantic Differencing for Activity Diagrams [*]

Shahar Maoz, Jan Oliver Ringert, Bernhard Rumpe
Software Engineering
RWTH Aachen University, Germany
http://www.se-rwth.de/

## ABSTRACT

Activity diagrams (ADs) have recently become widely used in the modeling of workflows, business processes, and web-services, where they serve various purposes, from documentation, requirement definitions, and test case specifications, to simulation and code generation. As models, programs, and systems evolve over time, understanding changes and their impact is an important challenge, which has attracted much research efforts in recent years.

In this paper we present *addiff*, a semantic differencing operator for ADs. Unlike most existing approaches to model comparison, which compare the concrete or the abstract syntax of two given diagrams and output a list of syntactical changes or edit operations, *addiff* considers the *semantics* of the diagrams at hand and outputs a set of *diff witnesses*, each of which is an execution trace that is possible in the first AD and is not possible in the second. We motivate the use of *addiff*, formally define it, and show two algorithms to compute it, a concrete forward-search algorithm and a symbolic fixpoint algorithm, implemented using BDDs and integrated into the Eclipse IDE. Empirical results and examples demonstrate the feasibility and unique contribution of *addiff* to the state-of-the-art in version comparison and evolution analysis.

## Categories and Subject Descriptors

D.2.2 [**Software Engineering**]: Design Tools and Techniques; D.2.4 [**Software Engineering**]: Software/Program Verification

[*]S. Maoz acknowledges support from a postdoctoral Minerva Fellowship, funded by the German Federal Ministry for Education and Research. J.O. Ringert is supported by the DFG GK/1298 AlgoSyn.

## 1. INTRODUCTION

Activity diagrams (ADs) have recently become widely used in the modeling of workflows, business processes, and web-services, where they serve various purposes, from documentation, requirement definitions, and test case specifications, to simulation and code generation. Specifically, we are interested in a variant of the standard UML 2 ADs, which is rich and expressive, supporting guarded branches, parallel (interleaving) process executions, inputs, assignments, etc.

As models, programs, and systems evolve over time, during the development lifecycle and beyond it, effective change management and controlled evolution are major challenges in software development, and thus have attracted much research efforts in recent years (see, e.g., [1, 2, 6, 11, 14, 21, 23, 32]). Fundamental building blocks for tracking the evolution of software artifacts are diff operators one can use to compare two versions of a program or a model. Most existing approaches to differencing concentrate on matching between model elements using different heuristics related to their names and structure and on finding and presenting differences at a concrete or abstract syntactic level. While showing some success, most of these approaches are also limited. Models that are syntactically very similar may induce very different semantics (in the sense of 'meaning' [10]), and vice versa, models that semantically describe the same system may have rather different syntactic representations. Thus, a list of syntactic differences, although accurate, correct, and complete, may not be able to reveal the real implications these differences may have on the correctness and potential use of the models involved. In other words, such a list, although easy to follow, understand, and manipulate (e.g., for merging), may not be able to expose and represent the semantic differences between two versions of a model, in terms of the bugs that were fixed or the features (and new bugs...) that were added.

In this paper we present *addiff*, a semantic diff operator for ADs. Unlike existing differencing approaches, *addiff* is a *semantic diff operator*. Rather than comparing the concrete or the abstract syntax of two given diagrams, and outputing a list of syntactical changes or edit operations, *addiff* considers the semantics of the diagrams at hand and outputs a set of *diff witnesses*, each of which is an execution trace that is possible in the first AD and is not possible in the second. These traces provide concrete proofs for the meaning of the change that has been done between the two compared versions and for its effect on the use of the models at hand.

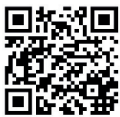

We specify ADs using a variant of standard UML 2 ADs [24], which can also be given textually using a grammar defined in MontiCore [13, 22]. The syntax of an AD consists of action nodes, pseudo nodes (initial, final, decision, merge, fork, join), transitions, input variables, and local variables. Transitions outgoing decision nodes are guarded with Boolean expressions over the input and local variables. Input variables values are set by the environment. Action nodes are labeled with action names and may include assignments to local variables.

We define the operational semantics of an AD using a translation to a finite automaton with variables over finite domains. This induces a trace-based semantics, i.e., a set of action traces from an initial node to a final node, considering also the values of input variables. Branches outgoing fork nodes describe parallel executions; they are used to succinctly specify concurrent interleaving traces. An overview of the syntax and semantics of our ADs is given in Sect. 3.

Given two ADs, $ad_1$ and $ad_2$, $addiff(ad_1, ad_2)$ is roughly defined as the set of execution traces possible in the first AD and not possible in the second. As there may be exponentially many diff traces, we are specifically interested in the shortest ones, i.e., ones that do not have a prefix which is a differentiating trace. In addition, we restrict the operator to provide only a single shortest diff trace for each possible assignment to input variables. To compute $addiff$ we transform each of the ADs into a module in SMV, the input language of the SMV model checker [29]. We then present two algorithms: a concrete forward-search algorithm and a symbolic fixpoint algorithm. The second algorithm relies on the technologies of symbolic model checking [3] in order to address the state explosion problem of the first. We present them both in order to allow their comparison. The formal definition of $addiff$ and the two algorithms are described in Sect. 4.

We have implemented the two algorithms for $addiff$ using binary decision diagrams (BDDs), and integrated them into an Eclipse plug-in. The plug-in allows the engineer to compare two selected ADs, to check if they are equivalent, and to textually and visually browse the diff witnesses found, if any. We describe the plug-in implementation in Sect. 5. We have evaluated the plug-in against all examples shown in this paper and many other ADs. The results of our evaluation, including a performance comparison of the two algorithms, appear in Sect. 6.

In addition to finding concrete diff witnesses (if any exist), which demonstrate the meaning of the changes that were made between one version and another, $addiff$ can be used to compare two ADs and decide whether one AD semantics includes the other AD semantics (the latter is a refinement of the former), are they semantically equivalent, or are they semantically incomparable (each allows executions that the other does not allow). When applied to the version history of a certain AD, as can be retrieved from a version repository, such an analysis provides a semantic insight into its evolution, which is not available in existing syntactic approaches.

Model and program differencing, in the context of software evolution, has attracted much research efforts in recent years. In contrast to our work, however, most studies in this area present syntactic differencing, at either the concrete or the abstract syntax level. We discuss related work in Sect. 7.

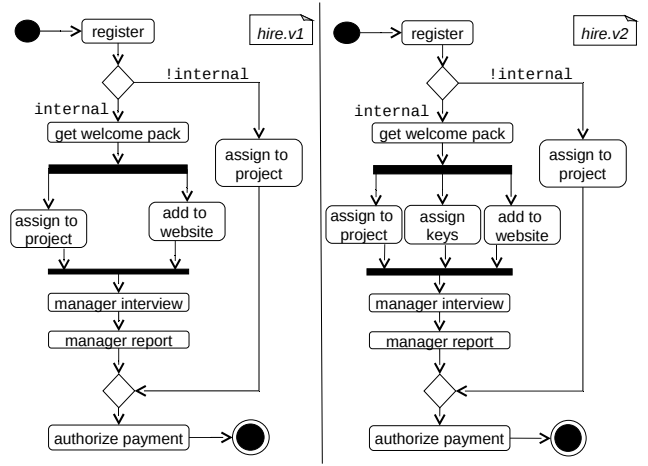

**Figure 1: Versions 1 and 2 of the hire activity**

Finally, our work on semantic differencing does not come to replace existing syntactic differencing approaches. Rather, it is aimed at augmenting and complementing existing approaches with capabilities that were not available before. We discuss the combination of syntactic and semantic differencing as well as other future work directions in Sect. 6.

The next section presents motivating examples demonstrating the unique features of our work. Sect. 3 provides preliminary definitions of the AD language syntax and semantics as used in our work. Sect. 4 introduces $addiff$ and the two algorithms to compute it. Sect. 5 presents the implementation, Sect. 6 presents an evaluation and a discussion, Sect. 7 considers related work, and Sect. 8 concludes.

## 2. EXAMPLES

We start off with motivating examples for semantic differencing of ADs. The examples are inspired by real-world ADs we have obtained from several sources (see Sect. 6).

## 2.1 Example I

AD $hire.v1$ of Fig. 1 describes a company's workflow when hiring a new employee. Roughly, first the employee is registered. Then, if she is an internal employee, she gets a welcome package, she is assigned to a project and added to the company's computer system (in two parallel activities branching off a fork node), she is interviewed and gets a manager report, and finally her payments are authorized. Otherwise (note the decision node at the beginning of the AD), if the new employee is external, she is only assigned to a project before her payments are authorized.

After some time, the company deployed a new security system and every employee had to receive a key card. A revised workflow was created, as shown in $hire.v2$ of Fig. 1.

Later, a problem was found: sometimes employees are assigned to a project but cannot enter the building since they do not have a key card yet. This bug was fixed in the next version, $hire.v3$, shown in Fig. 2. Finally, the company has decided that external employees should report to managers too. Thus, the merge between the two branches for internal and external new employees has moved 'up', in between the

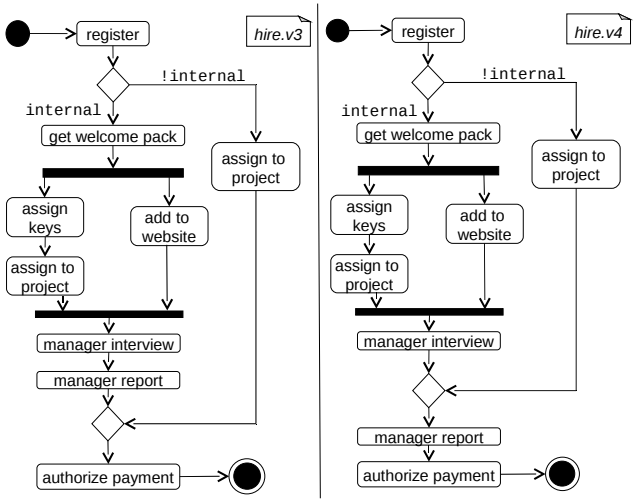

**Figure 2: Versions 3 and 4 of the hire activity**

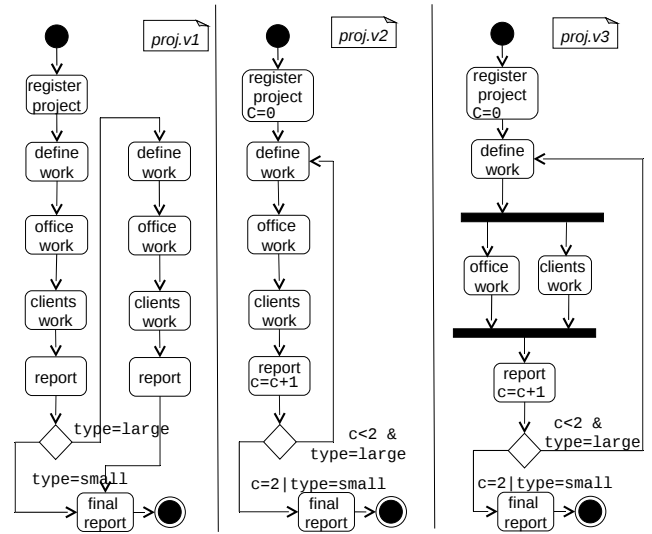

**Figure 3: Versions 1-3 of the new project activity**

interview and the report nodes. The resulting 4th version of the workflow, $hire.v4$, is shown in Fig. 2

Given these four versions of the activity, an evolution analysis is called for. Comparing $hire.v1$ and $hire.v2$ using $addiff$ reveals that they are semantically incomparable: some executions of $hire.v1$ are no longer possible in $hire.v2$, and some executions of $hire.v2$ were not possible in $hire.v1$. Moreover, it reveals that handling of internal employees has changed, but handling of external ones remained the same between the two versions.

Comparing $hire.v2$ and $hire.v3$ reveals that the latter is a refinement of the former: $hire.v3$ has removed some traces of $hire.v2$ and did not allow new traces. In particular, $addiff(hire.v2, hire.v3)$ shows that the trace where a person is assigned to a project before she gets a key card was possible in $hire.v2$ and is no longer possible in $hire.v3$, i.e., it demonstrates that the bug was fixed.

Finally, comparing $hire.v3$ and $hire.v4$ using $addiff$ reveals that although hiring of external employees has changed between the two versions, hiring of internal employees did not: $addiff(hire.v3, hire.v4)$ contains a single trace, where the employee is external, not internal. That is despite the syntactic change of moving the merge node from after to before the report node, which is also part of the trace of handling internal employees.

## 2.2 Example II

AD $proj.v1$ of Fig. 3 describes a company's workflow when receiving a new project. Roughly, first the project is registered. Then, the required work is defined, the office work is done, the clients work is done, and a report is written. If the project is of type small, the activity continues to final report and is completed. If the project is of type large, a second phase of the required work is defined, the office work is done, the clients work is done, and a report is written, before continuing to the final report and completion.

After some time, the activity designer suggested a refactoring: instead of the explicit duplication of the four work actions in the diagram, a loop will be defined. The designer added a local variable $c$, which is initialized when the project is registered and is incremented when writing the report.

The guards of the decision node were changed accordingly. The resulting AD is $proj.v2$, shown in the same figure.

Before committing the new activity to the models repository, the designer used $addiff$ to compare $proj.v1$ and $proj.v2$. Indeed, she found that $addiff(proj.v1, proj.v2) = addiff(proj.v2, proj.v1) = \emptyset$. This proves that the refactoring did not change the semantics of the activity and the new version can be safely committed to the repository.

Finally, a consultant has examined the new project activity and suggested that in some cases, clients work can be done before the office work. Following this recommendation, the designer added fork and join nodes to the activity, to define the office work action and the clients work action on separate branches of a fork. The resulting AD is $proj.v3$, shown in the same figure.

Comparing $proj.v2$ and $proj.v3$ using $addiff$ revealed that the new version has introduced several new traces that were not possible before: traces where clients work is done before office work (in the first or in the second iteration of the loop). Thus, the comparison demonstrated that the required enhancement was added. Moreover, the comparison showed that all traces of $proj.v2$ are still possible in $proj.v3$ (because $addiff(proj.v2, proj.v3) = \emptyset$). Thus, it proved that no behavior was lost.

The above examples are simple and thus immediately reveal the differences when looking at them. We use them in order to demonstrate our ideas. However, we have also done experiments with larger, synthetic and real-world AD's, where differences were manually much harder to find. See Sect. 6.

## 3. PRELIMINARIES

We define the AD language syntax and semantics as used in our work.

## 3.1 AD Language Syntax

An Activity Diagram is a structure
$AD = \langle A, V^{inp}, V^{loc}, AN, PN, T \rangle$ where:

- $A$ is a set of action names.

- $V^{inp}$ is a (possibly empty) set of immutable input variables over finite domains.
- $V^{loc}$ is a (possibly empty) set of local variables over finite domains.
- $AN$ is a set of action nodes $an_1, \ldots, an_k$. Each action node $an$ is labeled with an action name $acname(an) = ac \in A$, and a (possibly empty) set of assignment expressions to the variables in $V^{loc}$.
- $PN$ is a set of pseudo nodes, consisting of initial nodes $PN^{init}$, final nodes $PN^{fin}$, decision nodes $PM^{dec}$, merge nodes $PN^{mer}$, fork nodes $PN^{fork}$, and join nodes $PN^{join}$.
- $T$ is a set of transitions of the form $t = \langle n_{src}, n_{trg}, guard \rangle$ where $n_{src}, n_{trg} \in (AN \cup PN)$ and $guard$ is a Boolean expression over the variables in $V^{inp} \cup V^{loc}$. Unless $n_{src}$ is a decision node, $guard = \texttt{true}$.

We do not formally capture here obvious well-formedness rules and context conditions such as: initial nodes have no incoming transitions, final nodes have no outgoing transitions, fork nodes must be followed by join nodes to remove all concurrency when reaching a final node, actions should not repeat in different forked branches, etc.

In addition, we assume that the Boolean expressions used as guards on transitions outgoing decision nodes are semantically exclusive, that is, no assignment to the diagram variables makes more than one of them true. Thus, the input variables provide external non-determinism, while except for forked branches, our ADs are internally deterministic.

We consider two concrete syntax definitions for ADs: a concrete visual syntax based on UML 2 ADs and a concrete textual syntax defined using MontiCore [13, 22]. We omit the concrete syntax definitions from this paper.

The implementation we present in Sect. 5 supports a subset of full UML 2 ADs: it supports action nodes and pseudo nodes: initial, final, fork, join, decision and merge; each AD has exactly one initial node but may have multiple final nodes. For simplicity reasons but without loss of generality, in our implementation no two pseudo nodes can follow each other directly. In addition, local and input variables have to be declared as such in the first action node; all variables' finite domains need to be given as SMV types or enumerations; and each local variable needs to be assigned a value in the first action node. We support Boolean guards specified in the rich SMV expression language [4]. Assignments to local variables can be made from any action node using values from SMV expressions.

### 3.2 AD Language Semantics

We distinguish operational semantics and trace-based semantics. The operational semantics is based on the definition of a state machine step, taking the AD from one state to another, where a state consists of a set of current action nodes and an assignment to all input and local variables.

The main idea of our operational semantics is to conceptually translate each $ad \in AD$ into a finite state machine (FSM). Each state is a configuration containing the values of all local variables and input variables (recall that variables range over finite domains), the executed action, and some extra variables with information on the control flow of the AD. Based on this state and evaluated guards possible transitions are computed that lead to the next state with an executed action, possibly changed local variables values, and a new configuration of the control flow tracking variables.

We formally define the operational semantics using a transformation of an AD to a module of SMV, the language of the SMV model checker [20, 29]. Our translation is inspired by the work of [7], but extends this previous translation with support for data. The complete translation, together with a detailed example, appears in a separate document [17].

Local variables assist the AD engineer together with guards to control the execution sequences of actions, e.g., by defining loops or activating/disabling branches of decision nodes. Input variables, in contrast to local variables, are initialized by the environment and do not change during the run of an AD. The runs of two ADs are compared with the same input, i.e., where common input variables of both ADs have the same values. We consider all possible input values of both ADs when comparing them.

Our semantics of ADs is rather expressive: it considers values of input variables to be set by the environment (external non-determinism). However, we only support internal non-determinism through interleaved execution of forked branches, and not through non-deterministic decision nodes; the current configuration and the next executed action determine the next configuration. Thus, for each assignment to the input variables there could be many possible executions due to the interleaving semantics of fork nodes.

We define a trace-based semantics that is induced by the operational semantics. Traces are sequences of states from the state space of the AD's FSM.

DEFINITION 1 (AD STATE). *A state of an AD ad is an assignment to all variables defined in the* `VAR` *section of its SMV module. This includes the last executed node* `acnode` *and its action name* `ac`*, the values of variables $v \in ad.V^{inp} \cup ad.V^{loc}$ and the values of control flow tracking variables.*

AD states have a finite number of possible successor states that can be reached within one step. By construction, each such step executes an action or reaches a final node. We define the successors of an AD state $s$ as $successors(s)$.

DEFINITION 2 (SUCCESSOR STATES). *For every AD state $s$, $successors(s)$ is the set of AD states reachable from $s$ in one step of the SMV module.*

Each run of an AD starts with an initial state in the initial node. A sequence of successor states that describe a legal execution of the AD's FSM is a trace. A trace from the initial to a final node of the AD is an accepting trace.

DEFINITION 3 (AD TRACES). *A sequence of AD states $tr = s_0, s_1, \ldots, s_k$ of AD ad with $s_{i+1} \in successors(s_i)$ and $s_0.acnode \in ad.PN^{init}$ is called a trace. The set of all traces of an AD ad is denoted by $traces(ad)$. A trace is called an accepted trace if its last state's node is a final node of the AD. The set of all accepted traces of an AD ad is denoted by $acceptedTraces(ad)$.*

## 4. ADDIFF

### 4.1 Definitions

Given two AD states, $s_1 \in ad_1$ and $s_2 \in ad_2$, we say that $s_1$ and $s_2$ are *corresponding*, iff the action names and values of equally named input variables of the two states are the same. Formally:

DEFINITION 4 (CORRESPONDING STATES). *Given two ADs, $ad_1$ and $ad_2$, and two states $s_1 \in ad_1$ and $s_2 \in ad_2$, we say that $s_1$ and $s_2$ are corresponding, denoted $s_1 \sim s_2$, iff the following conditions hold:*

1. $s_1.ac = s_2.ac$;
2. $\forall v \in V_1^{inp} \cap V_2^{inp}$, $s_1.val(v) = s_2.val(v)$.

The definition of corresponding states punctually extends to traces in a natural way.

Given two ADs, a trace of the first AD is a *diff trace* iff there exists a trace of the second AD where all states except the last correspond to the states of the first trace, but the last state of the first trace does not correspond to any possible successor state of the second trace. Formally:

DEFINITION 5 (DIFF TRACE). *Given two ADs, $ad_1$ and $ad_2$, a diff trace is a sequence of states $tr_1 = s_1^0, s_1^1, \ldots, s_1^k, s_1^{k+1}$ s.t.*

1. $tr_1 \in traces(ad_1)$
2. $\exists tr_2 = s_2^0, s_2^1, \ldots, s_2^k$ s.t. $tr_2 \in traces(ad_2)$
   $\wedge \forall i, 0 \leq i \leq k, s_1^i \sim s_2^i$
   $\wedge \nexists s_2^{k+1}$ s.t. $s_2^{k+1} \sim s_1^{k+1} \wedge s_2^0, s_2^1, \ldots, s_2^k, s_2^{k+1} \in traces(ad_2)$.

$tr_2$ *is called a corresponding diff trace of $tr_1$.*

We denote the set of all diff traces of $ad_1$ vs. $ad_2$ by $diffTraces(ad_1, ad_2)$. Note that $diffTraces$ is not symmetric.

We are now ready to present the definition of *addiff*. Note that we are interested only in shortest diff traces: we restrict *addiff* to diff traces that do not have another diff trace as prefix. Moreover, to make the set *addiff* succinct, for each initial state of $ad_1$, if there is a diff trace that starts at this initial state, we want only one such trace to be in *addiff*. Formally:

DEFINITION 6 (ADDIFF). *$addiff(ad_1, ad_2)$ is a subset of $diffTraces(ad_1, ad_2)$ s.t.*

1. $\forall tr \in addiff(ad_1, ad_2), \nexists tr'$ s.t. $tr' \in addiff(ad_1, ad_2) \wedge tr' \sqsubset tr$;
2. $\forall s_1^0 \in ad_1.initials$,
   if $\exists tr \in diffTraces(ad_1, ad_2)$ s.t. $tr$ starts at $s_1^0$
   then $|\{tr \in addiff(ad_1, ad_2) | tr$ starts at $s_1^0\}| = 1$.

Finally, note that we do *not* require that diff traces can be extended into accepting traces (ones that end at a final node, see Def. 3). For an alternative definition, see the discussion in Sect. 6.

## 4.2 Computing ADDiff

We present two different algorithms, a concrete forward-search algorithm and a symbolic fixpoint algorithm.

### 4.2.1 Algorithm I

We compute $addiff(ad_1, ad_2)$ using a BFS-like traversal of the state space of $ad_1$ that is used to 'guide' a BFS-like traversal of the state space of $ad_2$. Roughly, the algorithm uses a queue for corresponding states-pairs that have been reached but whose successors have not yet been traversed (the use of the queue guarantees that shortest paths will be found first). It also maintains a list of visited corresponding state-pairs and a list of rejecting state-pairs. Initially, all corresponding initial states-pairs are inserted in the queue, and all initial states of $ad_1$ that do not have a corresponding initial state in $ad_2$ are added to the list of rejecting state-pairs. Then, for each state-pair taken out from the queue, the algorithm checks that each successor state of the first element in the state-pair (the state in $ad_1$), has a corresponding successor state of the second element in the state-pair (the state in $ad_2$). Every corresponding pair found, if not visited before, is inserted to the queue. If no corresponding successor is found, we know we have found the end state of a (shortest) diff trace: we add it to the list of rejecting state-pairs and we remove from the queue all the state-pairs whose input variables values for $ad_1$ are the same as the ones for the state we have found. When the queue is empty, the list of visited state-pairs is used to construct the traces leading back from the rejecting state-pairs to the initial states.

A pseudo-code for the algorithm is given in Proc. 1, which uses Proc. 2 and 3. We describe these procedures below.

The algorithm uses a structure *Pair* made of two pairs of states: predecessor and current state in $ad_1$, $pre_1$ and $cur_1$, and predecessor and current state in $ad_2$, $pre_2$ and $cur_2$. Two pairs are considered equal if their current states are equal (ignoring predecessor states). *Pair* is used to keep track of pairs of visited states, one state from each AD, and of their predecessors, as found during the traversal of the state space. The predecessors are used in the reconstruction of the traces from the lists of rejecting and visited pairs.

Proc. 1 defines the required structures (l. 1-3): a queue of pairs, a list of visited pairs, a list of rejecting pairs, and a list of list of pairs, which will hold the computed diff traces. It iterates over all initial states of $ad_1$, and for each of them, looks for a corresponding initial state in $ad_2$. If a corresponding state is found, the new pair is inserted to the queue and to the visited list (l. 8). If no corresponding state is found, a pair where only the $ad_1$ current state is defined is added to the list of rejecting state pairs (l. 13).

After initialization, the algorithm calls $traverse$ (Proc. 2), to iterate on the queue until it is empty. For each dequeued pair, the procedure iterates over all the successors of its current $ad_1$ state. For each successor, it tries to find a corresponding successor of the current $ad_2$ state. If a corresponding state is found, the new pair is inserted to the queue and to the visited list (l. 9). If no corresponding state is found, a pair where only the $ad_1$ current state is defined is added to the list of rejecting states (l. 15). In addition, all state-pairs whose current $ad_1$ state agrees with the current $ad_1$ state in the rejecting pair on input variables, are removed from the queue (l. 16-17). This ensures that no further searching of diff traces outgoing the same initial state will be done.

Finally, $trace$ (Proc. 3) is used to reconstruct the traces leading from initial states to the rejecting states that have been found. The procedure works backward: it starts from the rejecting pairs and uses the predecessor states to build the required traces, from the rejecting states back to the initial states, using the pairs saved in the *visited* list. It continues as long as their predecessor states are defined, i.e., as long as it has not reached an initial state.

### 4.2.2 Algorithm II

We compute $addiff(ad_1, ad_2)$ using a symbolic least-fixpoint algorithm. The algorithm relies on the technologies of symbolic model checking [3], and is inspired by the classic fixpoint algorithm to compute a maximal simulation relation, and, more specifically, by the synthesis algorithm of [26],

**Procedure 1** concrete-addiff($ad_1$,$ad_2$)
1: **define** $queuePairs$ **as queue of** $Pair$
2: **define** $visited$, $rejects$ **as list of** $Pair$
3: **define** $traces$ **as list of lists of** $Pair$
4: **for all** $ini_1 \in ad_1.initials$ **do**
5:   $foundCorresponding \leftarrow$ **false**
6:   **for all** $ini_2 \in ad_2.initials$ **do**
7:     **if** $corresponding(ini_1, ini_2)$ **then**
8:       **add** $Pair(-,ini_1,-,ini_2)$ **to** $queuePairs$, $visited$
9:       $foundCorresponding \leftarrow$ **true**
10:     **end if**
11:   **end for**
12:   **if not** $foundCorresponding$ **then**
13:     **add** $Pair(-,ini_1,-,-)$ **to** $rejects$
14:   **end if**
15: **end for**
16: $visited, rejects \leftarrow$ traverse($ad_1$,$ad_2$)
17: $traces \leftarrow$ trace($visited$,$rejects$)
18: **return** $traces$

---

**Procedure 2** traverse($ad_1$,$ad_2$)
1: **while** $queuePairs$ **is not empty do**
2:   $p \leftarrow pair$ **from** $queuePairs$
3:   **for all** $suc_1 \in p.cur_1.successors$ **do**
4:     $foundCorresponding \leftarrow$ **false**
5:     **for all** $suc_2 \in p.cur_2.successors$ **do**
6:       **if** $corresponding(suc_1, suc_2)$ **then**
7:         $newPair \leftarrow Pair(p.cur_1, suc_1, p.cur_2, suc_2)$
8:         **if** $newPair \notin visited$ **then**
9:           **add** $newPair$ **to** $queuePairs$, $visited$
10:         **end if**
11:         $foundCorresponding \leftarrow$ **true**, **break**
12:       **end if**
13:     **end for**
14:     **if not** $foundCorresponding$ **then**
15:       **add** $Pair(p.cur_1, suc_1, p.cur_2, -)$ **to** $rejects$
16:       **remove all** $pair$ **from** $queuePairs$
17:         **where** $p.cur_1.inputs = pair.cur_1.inputs$
18:     **end if**
19:   **end for**
20: **end while**
21: **return** $visited, rejects$

---

**Procedure 3** trace($visited$,$rejects$)
1: **for all** $rejectingPair \in rejects$ **do**
2:   **define** $tr$ **as list of** $Pair$
3:   $curPair \leftarrow rejectingPair$
4:   **while** $curPair$ **is not null do**
5:     **add** $curPair$ **to** $tr$
6:     **if** $curPair.pred_1$ **is not null then**
7:       $curPair \leftarrow$
        $getVisited(curPair.pred_1, curPair.pred_2, visited)$
8:     **else**
9:       **break**
10:     **end if**
11:   **end while**
12:   **add** $tr$ **to** $traces$
13: **end for**
14: **return** $traces$

---

**Procedure 4** symbolic-addiff($ad_1$,$ad_2$)
1: **define** $traces$ **as list of lists of** $Set$
2: **define** $mem$ **as array of** $Set$
3: $mem \leftarrow$ least-fixpoint-with-mem($ad_1$,$ad_2$)
4: **if** $mem.last \cap initials \neq \emptyset$ **then**
5:   $traces \leftarrow$ build-traces-from-mem($ad_1$,$ad_2$,$mem$)
6: **end if**
7: **return** $traces$

---

where intermediate values from the fixpoint computation are used in the construction of a concrete winning strategy.

Roughly, our symbolic algorithm starts with a representation of all non-corresponding states. It then moves 'backward', and adds to the current set of states, states from which there exists a successor in $ad_1$ such that for all successors in $ad_2$, the resulting successor pair is in the current set of states. Most importantly, to help in the construction of diff traces later, at each step backward, the algorithm remembers the newly computed set of added states. The steps 'backward' continue until reaching a least fixpoint, that is, until no more states are added.

When the fixpoint is reached, the algorithm checks whether the last computed set (the fixpoint set) includes initial states. For each such initial state, if any, the algorithm uses the sets of states computed during the backward steps to move forward (from the minimal position it can start from) and construct shortest diff traces.

We present our algorithm in general set notation, with the set-operations of union, intersection, and complementation. In the pseudo code below, sets with no subscript are sets of states over the union of all variables from $ad_1$ and $ad_2$. For $i \in \{1, 2\}$, sets with subscript $i$ are sets of states over the variables of $ad_i$. The operator $S|_{ad_i}$ is used to restrict the variables of the set $S$ to the variables of $ad_i$ (all other variables are existentially quantified out). The operation `choose one` relates to choosing a single element from the relevant set (a single concrete assignment to the variables). The sets $corr$ and $initials$ are the set of corresponding states and the set of joint initial states, respectively. When intersecting a set $S_i$ over the variables of $ad_i$ with a set $S$ over the union of all variables from both ADs, the result is a set over the union of variables where the variables of $ad_i$ agree with their possible assignments in the set $S_i$.

In the implementation, the sets are represented using BDDs. A pseudo-code for the algorithm is given in Proc. 4, which uses Proc. 5 and 6.

Note that in the final iteration of the loop in Proc. 6, reached with $i = 1$, the assignment to $next_2$ is guaranteed to set $next_2 \leftarrow \emptyset$, because the first location in the memory array equals $\overline{corr}$. This guarantees that the last state in each diff trace assigns no values to the variables of $ad_2$: indeed, the last $ad_1$ state in the trace should have no corresponding $ad_2$ state in the trace.

If the user is interested only in checking the existence of differences but not in the set of all witnesses, we can stop the steps backward as soon as the set of added states includes an initial state (by checking whether $p \cap initials = \emptyset$ already after line 9 in Proc. 5). In some cases, as our evaluation shows (see Sect. 6), this is indeed much faster than waiting for the fixpoint to be reached and for all traces to be enumerated. Note that trace enumeration is not symbolic and thus may not scale well.

Finally, for both algorithms, the concrete and the symbolic, correctness and completeness are proved by induction on the length of the traces and rely on the fact that the ADs are internally deterministic.

## 5. IMPLEMENTATION AND USES

We have implemented *addiff* and integrated it into a prototype Eclipse plug-in. The input for the implementation are

**Procedure 5** least-fixpoint-with-mem($ad_1$,$ad_2$)
1: **define** $mem$ **as** array of $Set$
2: **define** $p, z, oldz$ **as** $Set$
3: **define** $i$ **as** number
4: $z \leftarrow \overline{corr}, oldz \leftarrow \emptyset$
5: $i \leftarrow 0$
6: $mem[i] \leftarrow z$
7: **while** $z \neq oldz$ **do**
8: $\quad oldz \leftarrow z$
9: $\quad p \leftarrow \{ (s_1, s_2) \mid s_1 \in ad_1 \wedge s_2 \in ad_2 \wedge \exists suc_1 \in s_1.successors$
$\quad\quad$ s.t. $\forall suc_2 \in s_2.successors \ (suc_1, suc_2) \in z\}$
10: $\quad z \leftarrow z \cup p$
11: $\quad i \leftarrow i + 1$
12: $\quad mem[i] \leftarrow z$
13: **end while**
14: **return** $mem$

---

**Procedure 6** build-traces-from-mem($ad_1$,$ad_2$,$mem$)
1: **for all** $ini_1 \in (mem.last \cap initials)|_{ad_1}$ **do**
2: $\quad$ **define** $tr$ **as** list of $Set$
3: $\quad$ **find minimal** $j$ s.t. $ini_1 \cap mem[j] \neq \emptyset$
4: $\quad ini_2 \leftarrow$ **choose one from** $(ini_1 \cap initials)|_{ad_2}$
5: $\quad CS \leftarrow CombinedState(ini_1, ini_2)$
6: $\quad$ **add** $CS$ **to** $tr$
7: $\quad$ **for** $i = j$ **down to** $1$ **do**
8: $\quad\quad next_1 \leftarrow$ **choose one from**
$\quad\quad\quad (CS|_{ad_1}.successors \cap mem[i-1]|_{ad_1})$
9: $\quad\quad next_2 \leftarrow$ **choose one from**
$\quad\quad\quad (next_1 \cap corr \cap mem[i-1])|_{ad_2} \cap (CS|_{ad_2}.successors)$
10: $\quad\quad CS \leftarrow CombinedState(next_1, next_2)$
11: $\quad\quad$ **add** $CS$ **to** $tr$
12: $\quad$ **end for**
13: $\quad$ **add** $tr$ **to** $traces$
14: **end for**
15: **return** $traces$

---

UML 2 ADs, drawn and parsed using Eclipse UML 2 APIs. The plug-in transforms the input ADs into SMV format. It then computes *addiff* via the APIs of JTLV [27], a framework for the development of verification algorithms, using bdd-based symbolic mechanisms. The underlying BDD package used is CUDD [30]. Both algorithms are implemented and the engineer can choose which one to use. The plug-in, together with all the example ADs we used in the evaluation, is available from [28].

### 5.1 Browsing diff traces

The plug-in allows the engineer to compare two selected ADs, and to textually and visually browse the diff traces found, if any. Fig. 4 shows an example screenshot, where the engineer has selected to compare diagrams $hire.v2$ (top) and $hire.v4$ (middle) (presented in Sect. 2), and is browsing one of the two diff traces that were found. Note the numbered and highlighted action nodes, which visually show the states along one of the diff traces that the plug-in has found: the trace `register`, `get welcome pack`, `assign to project` is possible in $hire.v2$ and is not possible in $hire.v4$.

A textual representation of the diff trace is displayed on the lower pane. This representation of a trace is a special case of a *model-based trace* (see [15, 16]). It shows the details of each state in the current diff trace in both ADs, consisting of the action name and all variable values. This textual representation is important because it is more detailed and because it scales better than the visual representation when handling long traces.

Clicking `Check Difference` checks whether the semantics of the second AD includes the semantics of the first. Clicking `Compute Witnesses` computes the diff traces and shows a message telling the engineer how many traces were found, if any. The `Next` and `Previous` buttons browse for the next and previous diff traces. The `Switch Direction` button switches the order of comparison. The `Concrete` and `Symbolic` buttons toggle between the two *addiff* algorithms.

### 5.2 High-level evolution analysis

Another application enabled by the plug-in is high-level evolution analysis. The plug-in supports a `compare` command: given two ADs, $ad_1$ and $ad_2$, the command checks whether one AD is a refinement of the other, are the two ADs semantically equivalent, or are they semantically incomparable (each allows traces the other does not allow). Formally, $compare(ad_1, ad_2)$ returns one of four answers:

| | |
|---|---|
| $<$ | if $addiff(ad_1, ad_2) = \emptyset$ and $addiff(ad_2, ad_1) \neq \emptyset$ |
| $>$ | if $addiff(ad_1, ad_2) \neq \emptyset$ and $addiff(ad_2, ad_1) = \emptyset$ |
| $\equiv$ | if $addiff(ad_1, ad_2) = \emptyset$ and $addiff(ad_2, ad_1) = \emptyset$ |
| $<>$ | if $addiff(ad_1, ad_2) \neq \emptyset$ and $addiff(ad_2, ad_1) \neq \emptyset$ |

Given a reference to a series of historical versions of an AD, as can be retrieved from the AD's entry in a revision repository (e.g., SVN), the plug-in can use the `compare` command to compute a high-level analysis of the evolution of the AD: which new versions have introduced new behaviors relative to their predecessors, which new versions have eliminated behaviors relative to their predecessors, and which new versions included only syntactical changes that have not changed the semantics of the AD. For example, applying this evolution analysis to the ADs presented in Sect. 2 reveals: $hire.v1 <> hire.v2$, $hire.v2 > hire.v3$, $hire.v3 <> hire.v4$, $proj.v1 \equiv proj.v2$ and $proj.v2 < proj.v3$.

## 6. EVALUATION AND DISCUSSION

### 6.1 Evaluation

We have tested our implementations of *addiff* against synthetic ADs and against real-world ADs, selected and adapted from several sources: (1) selected ADs from a library of more than 700 business process models by IBM [8] (our selection is representative of the size and complexity statistics of the models in this library, as described in [8]), (2) several models (with version history) we have obtained from Nokia Corp., Test Management, and (3) from a third company (which requested to remain unnamed). The IBM models did not include version history information so we have manually added some mutations (e.g., action additions and removals, change of branching conditions). The models from Nokia and the third company included version history. All the models we have used are available in a dedicated evaluation project that is attached to the plug-in so that all experiments we report on below can be reproduced. The experiments were performed on a regular laptop computer, Intel Dual Core CPU, 2.8 GHz, with 4 GB RAM, running Windows Vista. Running times are reported in milliseconds.

#### 6.1.1 Quantitative evaluation

Table 1 (upper part) shows results from executing *addiff* over the examples presented in Sect. 2 and selected real-world ADs from the sources mentioned above. For each two versions we report the number of nodes, the size of the (reachable) state space, the number of diff traces found, the length of the shortest and longest diff traces found, and the times it took the concrete and the symbolic algorithms to (1) decide the existence of at least one diff trace and (2) to compute all diff traces.

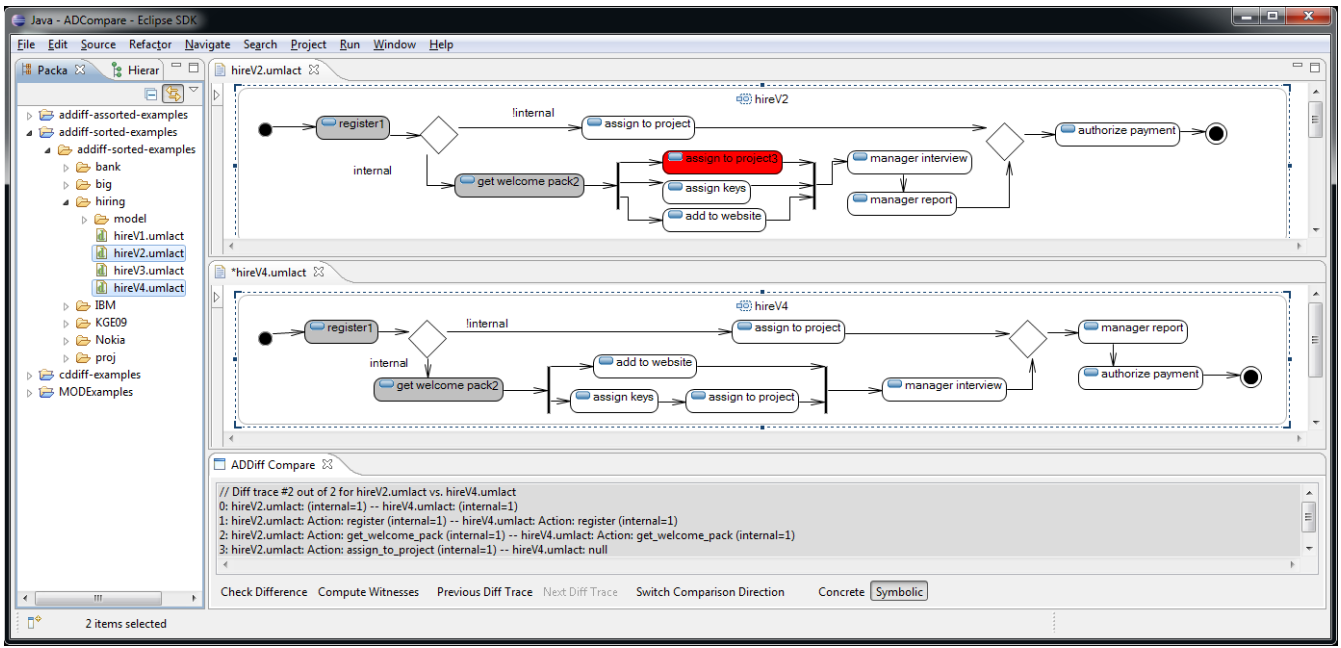

Figure 4: The prototype plug-in, comparing two ADs, $hire.v2$ and $hire.v4$. The highlighted and numbered nodes show one of the two diff traces found by $addiff$: for an internal employee, the trace `register`, `get welcome pack`, and `assign to project` is possible in $hire.v2$ and is not possible in $hire.v4$. A detailed textual representation of this diff trace is provided in the lower pane.

To examine scalability, we used synthetic ADs of comparable or much larger size (in both number of action nodes and state space), divided into two families of 'worst case' ADs: a family of 'forking ADs' with concurrent section of length $L$ and growing width $W$, and a family of 'linear ADs' with two linear fragments of length $L$ seperated by a single decision over a domain of increasing size $D$. The lower part of Table 1 shows results from executing $addiff$ on these ADs, with versions created using synthetic mutations: replacing, deleting, or moving of actions.

These results suggest the following observations. First, on small ADs, the two algorithms show similar performance results. However, as the ADs grow, the symbolic algorithm performs much better than the concrete one. On the largest ADs, the concrete algorithm performance is not practical while the symbolic algorithm stays within less than 4 seconds (on all real-world ADs). We believe this means $addiff$ can be used by engineers in practice.

Second, checking for the existence of a difference is sometimes much faster than listing all diff traces, specifically when there are many diff traces or when the shortest one is much shorter than the longest one. Again, as the ADs grow, the symbolic algorithm outperforms the concrete one.

That said, we do have synthetic 'worst case' examples where the symbolic algorithm is not better than the concrete one. This happens, e.g., in the extreme case of a long linear AD with no branches or forks. It also happens when the number of diff traces is large and enumerating them takes much time. Moreover, the specific change done between the two ADs may have significant effect on both algorithms' performance: two very syntactically similar mutations (e.g., a rename in one branch or in another), may induce dramatic changes in the number of diff traces. Complete results of and models used in our experiments are available from [28].

### 6.1.2 Qualitative evaluation

As mentioned above, we have obtained several real-world ADs with version history information from Nokia Corp., Test Management, and from another company. We used our plug-in to compare different versions and analyze the evolution of these ADs. We have also used other publicly available diff tools (Eclipse default differencing mechanism and EMFCompare [5]) in the analysis and compared them with $addiff$. Screen captures from our analysis sessions are available from [28].

Classical textual differencing was, as expected, not helpful in comparing the ADs, as it merely compared their XMI representations. EMFCompare [5] reported correct addition and removal of actions and transitions and presented them on the abstract syntax tree structures of the two ADs. Still, this was not so helpful in understanding the differences between the ADs because (1) the comparison was done on the AST level and abstracted the control flow away, and (2) the results were shown on the AST and not on the ADs themselves: we saw the changes in the AST and had to manually search for their concrete manifestation on the ADs. Moreover, in many cases, the list of additions and removals reported by EMFCompare was too large to be useful, e.g., when comparing $hire.v2$ and $hire.v4$ (presented earlier), EMFCompare reported 23 additions and removals, much more than one would intuitively expect for these two

| AD names | # Nodes | Reachable state spaces | # Wit. | Shortest/Longest | Alg. I decide/all (ms) | Alg. II decide/all (ms) |
|---|---|---|---|---|---|---|
| hireV1/hireV2 | 14/15 | 18/26 | 1 | 6/6 | 69/80 | 57/75 |
| hireV2/hireV3 | 15/15 | 26/21 | 1 | 4/4 | 53/69 | 53/68 |
| hireV3/hireV4 | 15/15 | 21/22 | 1 | 4/4 | 50/60 | 54/61 |
| projV1/projV2 | 13/9 | 22/22 | 0 | 0/0 | 52/52 | 44/47 |
| projV2/projV3 | 9/11 | 22/28 | 0 | 0/0 | 53/53 | 41/50 |
| IBM3561-1/2 | 18/18 | 121/122 | 2 | 7/61 | 105/173 | 113/243 |
| IBM2905-1/2 | 39/39 | 2680/2680 | 80 | 9/10 | 3010/7360 | 911/3667 |
| IBM2568-1/2 | 50/50 | 3834/3834 | 128 | 5/5 | 2725/4503 | 1328/3508 |
| IBM0863-1/2 | 23/23 | 1136/1136 | 76 | 5/8 | 1186/2970 | 200/844 |
| IBM3735-1/2 | 16/15 | 118/99 | 5 | 4/9 | 57/101 | 57/89 |
| IBM2557-1/2 | 17/15 | 275/175 | 6 | 7/7 | 66/95 | 74/120 |
| NokiaAV1/2 | 15/17 | 38/44 | 4 | 4/9 | 48/67 | 47/71 |
| NokiaAV2/3 | 17/23 | 44/96 | 4 | 7/7 | 71/101 | 76/111 |
| NokiaAV3/4 | 23/23 | 96/76 | 4 | 7/7 | 84/119 | 98/137 |
| AnonV1/2 | 15/15 | 37/37 | 2 | 6/6 | 50/64 | 47/61 |
| AnonV2/3 | 15/19 | 37/80 | 1 | 8/8 | 67/72 | 57/70 |
| forking(W1/L6)/mutated | 12/12 | 11/11 | 1 | 9/9 | 57/79 | 49/58 |
| forking(W2/L6)/mutated | 18/18 | 89/89 | 1 | 15/15 | 121/135 | 100/138 |
| forking(W3/L6)/mutated | 24/24 | 887/887 | 1 | 21/21 | 1293/1332 | 496/667 |
| forking(W4/L6)/mutated | 30/30 | 8237/8237 | 1 | 27/27 | 88878/89399 | 5892/7757 |
| lbl(L12/D16)/mutated | 34/34 | 496/496 | 1 | 15/15 | 367/608 | 373/507 |
| lbl(L12/D32)/mutated | 34/34 | 992/992 | 1 | 15/15 | 698/1377 | 678/895 |
| lbl(L12/D64)/mutated | 34/34 | 1984/1984 | 1 | 15/15 | 2242/4874 | 1651/2161 |
| lbl(L12/D128)/mutated | 34/34 | 3968/3968 | 1 | 15/15 | 9867/23529 | 5789/6685 |

Table 1: Results from computing *addiff* for selected example, real-world, and synthetic ADs (see Sect. 6.1.1). For each two versions we report the number of nodes, the reachable state space, the number of diff traces found, the length of the shortest and longest diff traces found, and the times it took the concrete and the symbolic algorithms to (1) decide the existence of at least one diff trace and (2) to compute all diff traces.

ADs (*addiff* reports a total of three diff witnesses for this example (two in one direction, one in the other)).

In contrast, our plug-in computed diff traces and highlighted them, visually, on the ADs themselves. Thus, in addition to the semantic characteristics of the comparison, which shows the actual meaning of the changes that were done, we have also experienced the advantages of language-specific differencing over language-agnostic differencing as well as the advantage of showing the differences directly on the original diagrams rather than in a separate list.

The following lessons learned are noteworthy. First, in some cases the number of traces returned by *addiff* was large and the usefulness of the results was limited. To address this in the future we consider adding filters, e.g., to group together traces that agree on the list of actions and differ in the values of input variables, and present only a representative trace from each group. Similarly, we consider user interaction: the engineer would choose a node of interest and the plug-in would limit the results to diff traces that include/exclude this node.

Second, some of the ADs we have analyzed included 'swim lanes', which relate action nodes with roles. It seems that 'swim lanes', which are optional in the UML standard [24], are rather popular, so in the future it may be useful to add the role information to the semantics of ADs and consider it in computing the differences.

Finally, some of the ADs we have investigated were only semi-formal or included minor changes in action names, which seem to indeed be 'renames' rather than new actions with similar names. *addiff* considers such 'renames' as new actions, and we had to manually identify these cases and 'correct' them. To better address these cases in the future, a matching heuristics needs to be employed, based perhaps not only on syntactic structural similarity matching but also on natural language and domain-specific ontology.

## 6.2 Discussion

### 6.2.1 Alternatives and extensions

Our current definition of diff traces does not require that they can be extended into accepting traces (ones that end at a final node, see Def. 3). We have chosen not to require this, in order to support the comparison of incomplete and perhaps inconsistent ADs, ones where not all executions are eventually accepted. Such ADs may exist, mainly at the early stages of the version history of a model (indeed in our evaluation we have seen such 'incomplete' ADs). We could have given a more restrictive definition that limits diff traces to ones that can be extended to accepted traces. Adapting the algorithms we have presented to this restricted definition is not difficult.

Moreover, we have chosen to compute only a single shortest representative of the diff traces outgoing each initial AD state (that is, a single shortest diff trace for each assignment to input variables of the first AD). We consider this to be a good choice, as it keeps the *addiff* results relatively succinct and thus easy to read and understand by engineers, in most typical cases. Alternatively, one may suggest to compute a larger set, containing all diff traces. Adapting the algorithms to this more permissive definition is possible, however, as there may be exponentially many such traces, performance may be a problem. On the other hand, and in contrast, following the lesson learned in our evaluation we consider an alternative that would limit the number of diff traces to present: group them according to the list of actions they include and present a single representative trace of each group together with a predicate that describes the input variable values that are common to the traces in the group. We believe this may be computed symbolically, i.e., while avoiding the enumeration of all traces in the computation. We leave this for future work.

Finally, our current work supports a subset of the UML 2 AD language. In particular, we do not yet support structured activity nodes, which allow hierarchical nesting of action nodes or reference from a node to another activity. Hierarchy is useful in medium and large scale designs, so supporting it is important. Moreover, a hierarchy induces an abstraction mechanism, which a semantic comparison may take advantage of. Additional language features may be added. We leave these for future work.

### 6.2.2 Syntactic differencing and matching

Semantic differencing in general, and *addiff* in particular, do not come to replace existing syntactic differencing approaches. Rather, they are aimed at augmenting and complementing existing approaches with capabilities that were not available before. Thus, combining *addiff* with existing approaches to matching and syntactic differencing (see, e.g., [14, 32]), is an important direction for future work. For example, we may extend the applicability of semantic differencing in comparing diagrams whose elements have been renamed or moved in the course of evolution, by applying a syntactic matching before running the semantic differencing. The result of such an integrated solution would be a mapping plus a set of diff traces. As another example, we may use information extracted from syntactic differencing as a means to localize and improve the presentation and performance of the semantic differencing computation.

## 7. RELATED WORK

We discuss related work on AD formal semantics and analyses, and on model and program comparisons.

Eshuis [7] presents symbolic model checking of ADs. The work transforms ADs into SMV and uses the NuSMV model checker to verify LTL properties. The semantics given is partly inspired by the semantics of STATEMATE [9]. Our translation of ADs into SMV is somewhat similar to the two translations suggested in [7]. [7] does not handle data while our work does. The motivation of [7] is model checking while our motivation is model comparison. Störrle [31] defines a denotational semantics for UML 2 ADs as a mapping to procedural Petri nets. He also surveys and compares several previous studies that deal with a semantics for ADs, in terms of their semantic domain and expressiveness. Knieke and Goltz [12] present an executable semantics for UML 2 ADs with step semantics adapted from [9]. The works of [7, 12] support object nodes and several types of action nodes, while our current work supports only basic action nodes. Our work can be extended to support object nodes and other types of action nodes. Our focus is not on the different possible variants of ADs and their semantics but on the definition and the computation of the semantic diff operator we have presented, and on its use in evolution related tasks.

Model and program differencing, in the context of software evolution, has attracted much research efforts in recent years (see [1, 6, 14, 21, 23, 32]). In contrast to our work, almost all studies in this area, however, present syntactic differencing, at either the concrete or the abstract syntax level.

Alanen and Porres [1] describe the difference between two models as a sequence of elementary transformations, such as element creation and deletion and link insertion and removal; when applied to the first model, the sequence of transformations yields the second. Kuster et al. [14] investigate differencing and merging in the context of process models, focusing on identifying dependencies and conflicts between change operations. Engel et al. [6] present the use of a model merging language to reconcile model differences. Comparison is done by identifying new/old MOF IDs and checking related attributes and references recursively. Results include a set of additions and deletions, highlighted in a Diff/Merge browser. Mehra et al. [21] describe a visual differentiation tool where changes are presented using editing events such as add/remove shape/connector etc. Xing and Stroulia [32] present an algorithm for object-oriented design differencing whose output is a tree of structural changes, reporting differences in terms of additions, deletions, and moves of model elements, assisted by a set of similarity metrics. Ohst et al. [23] compare UML documents by traversing their abstract-syntax trees, detecting additions, deletions, and shifts of sub-trees.

As the above shows, some works go beyond the concrete textual or visual representation and have defined the comparison at the abstract-syntax level, detecting additions, removals, and shifts operations on model elements. However, to the best of our knowledge, no previous work considers model comparisons at the level of the semantic domain, as is done in our work.

Some works, e.g. [5, 32], use similarity-based matching before actual differencing. As our work focuses on semantics, it assumes a matching is given. Matching algorithms may be used to suggest a matching before the application of semantic differencing. The result of such an integration would be a mapping plus a set of differentiating traces.

We are aware of only a few studies of semantic differencing between programs. Jackson and Ladd [11] summarise the semantic diff between two procedures in terms of observable input-output behaviors. Apiwattanapong et al. [2] present a behavioral diff for object-oriented programs based on an extended control-flow graph, and a tool that implements it in the context of Java. Finally, Person et al. [25] suggest to compute a behavioral characterization of a program change using a technique called differential symbolic execution. We focus on model comparison and not on program comparison. Also, while our work is somewhat similar to these works in terms of motivation, it is very different in terms of technology.

## 8. CONCLUSION

We presented *addiff*, a semantic differencing operator for activity diagrams. Unlike existing approaches to model comparison, *addiff* performs a semantic comparison and outputs a set of diff witnesses, each of which is an execution trace that is possible in the first AD and is not possible in the second. We have formally defined *addiff*, described two algorithms to compute it, a concrete one and a more scalable symbolic one, and demonstrated its application in comparing ADs within the Eclipse IDE. *addiff* can help developers to understand and evaluate the differences between versions of ADs so that they can reason about the impact of changes. When applied to the version history of a given AD, *addiff* provides a semantic insight into its evolution, which is not available in existing syntactic approaches.

We suggested a number of future work directions in Sect. 6, among them, the development of more succinct, symbolic, or task-oriented representation of diff traces, the integration of *addiff* with existing approaches to matching and syntactic differencing, and the extension of *addiff* to cover a larger subset of the UML 2 AD language, in particular, handling hierarchical actions, which are useful when considering the specification of medium and large scale activities.

Finally, *addiff* is part of a larger project that applies the idea of semantic differencing and computation of diff witnesses to several modeling languages [19]. We have recently presented our work on semantic differencing for class diagrams [18] and hope to report additional results from this project in future papers.

**Acknowledgements** We are grateful to Tuula Pääkkönen for help in obtaining the models from Nokia Corp., to Dirk Fahland for suggesting the use of the library of process models by IBM, to Yaniv Sa'ar for advice on the implementation of the algorithms in JTLV, and to Smadar Szekely and Guy Weiss for advice on Eclipse plug-in development.

## 9. REFERENCES

[1] M. Alanen and I. Porres. Difference and union of models. In P. Stevens, J. Whittle, and G. Booch, editors, *Proc. 6th Int. Conf. on the UML*, volume 2863 of *LNCS*, pages 2–17. Springer, 2003.

[2] T. Apiwattanapong, A. Orso, and M. J. Harrold. JDiff: A differencing technique and tool for object-oriented programs. *Autom. Softw. Eng.*, 14(1):3–36, 2007.

[3] J. R. Burch, E. M. Clarke, K. L. McMillan, D. L. Dill, and L. J. Hwang. Symbolic model checking: $10^{20}$ states and beyond. *Inf. Comput.*, 98(2):142–170, 1992.

[4] R. Cavada, A. Cimatti, C. A. Jochim, G. Keighren, E. Olivetti, M. Pistore, M. Roveri, and A. Tchaltsev. NuSMV User Manual, 2005.

[5] EMF Compare. http://www.eclipse.org/modeling/emft/?project=compare.

[6] K.-D. Engel, R. F. Paige, and D. S. Kolovos. Using a model merging language for reconciling model versions. In A. Rensink and J. Warmer, editors, *ECMDA-FA*, volume 4066 of *LNCS*, pages 143–157. Springer, 2006.

[7] R. Eshuis. Symbolic model checking of UML activity diagrams. *ACM Trans. Softw. Eng. Methodol.*, 15(1):1–38, 2006.

[8] D. Fahland, C. Favre, B. Jobstmann, J. Koehler, N. Lohmann, H. Völzer, and K. Wolf. Instantaneous soundness checking of industrial business process models. In U. Dayal, J. Eder, J. Koehler, and H. A. Reijers, editors, *BPM*, volume 5701 of *LNCS*, pages 278–293. Springer, 2009.

[9] D. Harel and A. Naamad. The STATEMATE Semantics of Statecharts. *ACM Trans. Softw. Eng. Methodol.*, 5(4):293–333, 1996.

[10] D. Harel and B. Rumpe. Meaningful modeling: What's the semantics of "semantics"? *IEEE Computer*, 37(10):64–72, 2004.

[11] D. Jackson and D. A. Ladd. Semantic diff: A tool for summarizing the effects of modifications. In H. A. Müller and M. Georges, editors, *ICSM*, pages 243–252. IEEE Computer Society, 1994.

[12] C. Knieke and U. Goltz. An executable semantics for UML 2 activity diagrams. In *Proc. Int. Workshop on Formalization of Modeling Languages (FML)*, 2010.

[13] H. Krahn, B. Rumpe, and S. Völkel. MontiCore: a framework for compositional development of domain specific languages. *Int. J. on Software Tools for Technology Transfer (STTT)*, 12(5):353–372, 2010.

[14] J. M. Küster, C. Gerth, and G. Engels. Dependent and conflicting change operations of process models. In R. F. Paige, A. Hartman, and A. Rensink, editors, *ECMDA-FA*, volume 5562 of *LNCS*, pages 158–173. Springer, 2009.

[15] S. Maoz. Model-based traces. In M. R. V. Chaudron, editor, *MoDELS Workshops*, volume 5421 of *LNCS*, pages 109–119. Springer, 2008.

[16] S. Maoz. Using model-based traces as runtime models. *IEEE Computer*, 42(10):28–36, 2009.

[17] S. Maoz, J. O. Ringert, and B. Rumpe. An Operational Semantics for Activity Diagrams using SMV. Technical Report AIB 2011-07, RWTH Aachen University, Germany, 2011.

[18] S. Maoz, J. O. Ringert, and B. Rumpe. CDDiff: Semantic differencing for class diagrams. In M. Mezini, editor, *Proc. 25th Euro. Conf. on Object Oriented Programming (ECOOP'11)*, volume 6813 of *LNCS*, pages 230–254. Springer, 2011.

[19] S. Maoz, J. O. Ringert, and B. Rumpe. A manifesto for semantic model differencing. In J. Dingel and A. Solberg, editors, *MoDELS Workshops*, volume 6627 of *LNCS*, pages 194–203. Springer, 2011.

[20] K. McMillan. *Symbolic Model Checking*. Kluwer Academic Publishers, 1993.

[21] A. Mehra, J. Grundy, and J. Hosking. A generic approach to supporting diagram differencing and merging for collaborative design. In *ASE*, pages 204–213. ACM, 2005.

[22] MontiCore project. http://www.monticore.org/.

[23] D. Ohst, M. Welle, and U. Kelter. Differences between versions of UML diagrams. In *Proc. ESEC / SIGSOFT FSE*, pages 227–236. ACM, 2003.

[24] OMG. UML, version 2.2, OMG Specification, 2009.

[25] S. Person, M. B. Dwyer, S. G. Elbaum, and C. S. Pasareanu. Differential symbolic execution. In *SIGSOFT FSE*, pages 226–237. ACM, 2008.

[26] N. Piterman, A. Pnueli, and Y. Sa'ar. Synthesis of reactive(1) designs. In E. A. Emerson and K. S. Namjoshi, editors, *VMCAI*, volume 3855 of *LNCS*, pages 364–380. Springer, 2006.

[27] A. Pnueli, Y. Sa'ar, and L. Zuck. JTLV: A framework for developing verification algorithms. In T. Touili, B. Cook, and P. Jackson, editors, *CAV*, volume 6174 of *LNCS*, pages 171–174. Springer, 2010.

[28] Semantic diff project. http://www.se-rwth.de/materials/semdiff/.

[29] SMV model checker. http://www.cs.cmu.edu/~modelcheck/smv.html.

[30] F. Somenzi. CUDD: CU Decision Diagram package. http://vlsi.colorado.edu/~fabio/CUDD/, 1998.

[31] H. Störrle. Semantics of control-flow in UML 2.0 activities. In *VL/HCC*, pages 235–242. IEEE Computer Society, 2004.

[32] Z. Xing and E. Stroulia. Differencing logical UML models. *Autom. Softw. Eng.*, 14(2):215–259, 2007.